\documentclass[prd,showpacs,preprintnumbers,amsmath,amssymb]{revtex4}

 \def\be{\begin{equation}}
 \def\ee{\end{equation}}
 \def\bes{\begin{eqnarray}}
 \def\ees{\end{eqnarray}}

 \def\p{\partial}
 
 \def\t{\tau}

 \def\2{\frac{1}{2}}
 \def\4{\frac{1}{4}}

\catcode`\@=11
\def\@citex[#1]#2{%
\if@filesw \immediate \write \@auxout {\string \citation {#2}}\fi
\@tempcntb\m@ne \let\@h@ld\relax \def\@citea{}%
\@cite{%
  \@for \@citeb:=#2\do {%
    \@ifundefined {b@\@citeb}%
      {\@h@ld\@citea\@tempcntb\m@ne{\bf ?}%
      \@warning {Citation `\@citeb ' on page \thepage \space
undefined}}%
      {\@tempcnta\@tempcntb \advance\@tempcnta\@ne%
      \@tempcntb\number\csname b@\@citeb \endcsname \relax%
      \ifnum\@tempcnta=\@tempcntb 
it
        \ifx\@h@ld\relax%
          \edef \@h@ld{\@citea\csname b@\@citeb\endcsname}%
        \else%
          \edef\@h@ld{\ifmmode{-}\else--\fi\csname
b@\@citeb\endcsname}%
        \fi%
      \else
        \@h@ld\@citea\csname b@\@citeb \endcsname%
        \let\@h@ld\relax%
      \fi}%
    \def\@citea{,\penalty\@highpenalty\,}%
  }\@h@ld
}{#1}}

\def\@citeb#1#2{{[#1]\if@tempswa , #2\fi}}
\def\@citeu#1#2{{$^{#1}$\if@tempswa , #2\fi }}
\def\@citep#1#2{{#1\if@tempswa , #2\fi}}

\begin{document}
\preprint{UTHET-08-1001}

\title{Dissipative Bjorken hydrodynamics from an AdS Schwarzschild black hole}

\author{James Alsup}
\email{jalsup1@utk.edu}
\author{George Siopsis}
 \email{siopsis@tennessee.edu}
\affiliation{%
Department of Physics and Astronomy,
The University of Tennessee,
Knoxville, TN 37996 - 1200, USA.
}%
\date{November 2008}%

\begin{abstract}

We discuss the derivation of dissipative Bjorken hydrodynamics from a Schwarzschild black hole in asymptotically AdS spacetime of arbitrary dimension in the limit of large longitudinal proper time $\tau$.
Using an appropriate slicing near the boundary, we calculate the Schwarzschild metric to next-to-next-to-leading order in the large $\tau$ expansion as well as the dual stress-energy tensor on the boundary via holographic renormalization. At next-to-next-to-leading order, it is necessary to perturb the Schwarzschild metric in order to maintain boost invariance.
The perturbation has a power law time dependence and leads to the same value of the ratio of viscosity to entropy density, $1/(4\pi)$, as in the case of sinusoidal perturbations. Our results are in agreement with known time-dependent asymptotic solutions of the Einstein equations in five dimensions.

 \end{abstract}

\maketitle

\section{Introduction}
\label{sec1}

There has been a substantial recent effort toward explaining the experimental results at the Relativistic Heavy Ion Collider (RHIC) in terms of a strongly coupled $\mathcal{N}=4$ Super Yang-Mills (SYM) gauge theory.  While the plasma formed at RHIC is described by a different gauge theory, namely Quantum Chromodynamics (QCD), one hopes that there are enough similarities between the two theories allowing us to aqcuire
a good qualitative (if not quantitative) understanding of the behavior of the quark-gluon plasma.
This will be achieved via the AdS/CFT correspondence \cite{adscft,adscftrev} which emerges in string theory and provides a method for studying the properties of strongly coupled $\mathcal{N}=4$ SYM theory in terms of a dual gravitational description.

The AdS/CFT correspondence has been applied to the description of the quark-gluon plasma in several different scenarios; see \cite{MV,Shuryak} for nice reviews.  Janik and Peschanski showed that a model for relativistic heavy ion collisions, suggested two decades previously by Bjorken \cite{Bjorken}, can be seen to be a consequence of the AdS/CFT correspondence \cite{JP}.  They obtained an approximate solution to the Einstein equations in the five-dimensional bulk in the large longitudinal proper time ($\tau$) limit and showed that it gave rise to hydrodynamics on the four-dimensional boundary matching the Bjorken flow of an ideal fluid \cite{Bjorken}.  The work was subsequently continued to include subleading terms in the large $\tau$ expansion which could be understood as dissipative effects in the hydrodynamic behavior of the gauge theory plasma \cite{NakSin,SinNakKim,ASM,Janik}.

Kajantie, {\em et al.,} \cite{KLT} showed that the two-dimensional Bjorken flow could be derived from a static black hole in the three-dimensional bulk spacetime by an appropriate time-dependent slicing near its boundary.
We extended this result to arbitrary dimensions \cite{AS} by showing that to leading order in $\tau$ there exists a slicing near the boundary of an AdS Schwarzschild black hole which corresponds to Bjorken flow in the dual gauge theory plasma. The Schwarzschild metric to leading order in $\tau$ agreed with the time-dependent asymptotic solution of Janik and Peschanski \cite{JP} in five dimensions. In three dimensions it reduced to the result of Kajantie, {\em et al.,} \cite{KLT}.

The aim of this paper is to extend our earlier result \cite{AS} by including subleading corrections in the large $\tau$ expansion.
We show that next-to-leading-order corrections correspond to viscosity in the gauge theory plasma. At this level the coefficient of viscosity $\eta$ is arbitrary, in agreement with results in five dimensions based on an asymptotic time-dependent solution to the Einstein equations \cite{NakSin}.
At next-to-next-to-leading order we find that the Schwarzschild metric yields a flow which is not boost-invariant no matter how one chooses the slicing near the AdS boundary. Boost invariance is recovered after the Schwarzschild metric is perturbed by a power law $\tau$-dependent perturbation. We show that the perturbed metric is non-singular in the bulk, provided
\be\label{eq1} \frac{\eta}{s} = \frac{1}{4\pi} \ee
where $s$ is the entropy density, in agreement with asymptotic time-dependent solutions in five dimensions \cite{Janik}.
This special value of the ratio $\eta/s$ is also in agreement with the case of sinusoidal perturbations of an AdS Schwarzschild black hole \cite{son}.


Our discussion starts with a review of dissipative Bjorken hydrodynamics in section \ref{sec2}.
In section \ref{sec3} we discuss the time dependent slicing we perform near the boundary in order to reproduce Bjorken hydrodynamics including next-to-next-to-leading-order contributions in the large $\tau$ expansion. We also introduce the perturbation to the Schwarzschild metric which is necessary to maintain boost invariance at the order we are interested in. We show that demanding absence of singularities in the bulk metric leads to the standard value (\ref{eq1}) of the viscosity to entropy density ratio. Finally section \ref{sec4} contains our concluding remarks.

\section{Dissipative Bjorken hydrodynamics}
\label{sec2}


Following Bjorken \cite{Bjorken}, in order to understand heavy ion collisions, we need to study boost invariant hydrodynamics.
Let us consider a gauge theory fluid on a $(d-1)$-dimensional flat Minkowski space spanned by coordinates $\tilde x^\mu$ ($\mu = 0,1,\dots,d-2$).
(We shall reserve the notation $x^\mu$ for the coordinates of a static gauge theory fluid; $\tilde x^\mu$ will span the Minkowski space of the Bjorken fluid in order to avoid confusion.)
With the colliding beams along the $\tilde x^1$ direction, it is convenient to work with the coordinates $\tau$ (longitudinal proper time) and $y$ (rapidity), defined by
\be \tilde x^0 = \tau \cosh y \ \ , \ \ \ \ \tilde x^1 = \tau \sinh y \ee
The $(d-1)$-dimensional Minkowski metric takes the form
\be\label{eq2} ds_{\mathrm{Bjorken}}^2 = d\tilde x_\mu d\tilde x^\mu = - d\tau^2 + \tau^2 dy^2 + (d\tilde x^\perp)^2 \ee
where $\tilde x^\perp = (\tilde x^2, \dots,\tilde x^{d-2})$ represents the tranverse coordinates.

For the stress-energy tensor we use the standard notions and enforce conservation and conformal invariance via:
\bes\label{Stensor}
T^{\mu\nu} &=& (\varepsilon+p)u^\mu u^\nu+p g^{\mu\nu}-\zeta\triangle^{\mu\nu}\nabla_\lambda u^\lambda -\eta\left(\triangle^{\mu\lambda}\nabla_\lambda u^\nu+\triangle^{\nu\lambda}\nabla_\lambda u^\mu-\frac{2}{d-2}\triangle^{\mu\nu}\nabla_\lambda u^\lambda\right)\nonumber\\
\nabla_\mu T^{\mu\nu} &=& 0\nonumber\\
T_\mu^\mu &=& 0
\ees
where $\triangle_{\mu\nu}=g_{\mu\nu}+u_\mu u_\nu$ and $\varepsilon$, $p$, $\eta$ and $\zeta$ represent the energy density, pressure, shear viscosity and bulk viscosity, respectively, of the fluid.
Two constraints immediately follow from eq.~(\ref{Stensor}),
\be\label{eqpz} \varepsilon=(d-2)p\ \ , \ \ \ \ \zeta=0\ee
In the rest frame of the conformal fluid, the velocity field is
given by $u^\mu=(1,\vec{0})$.
The stress-energy tensor simplifies to
%
\be\label{Vstensor}
T^{\mu\nu} =
\left(\begin{array}{cccc}
\varepsilon (\t) & 0 & \dots & 0 \\
0 & \frac{p(\t)}{\tau^2}-2\frac{d-3}{d-2}\frac{\eta(\tau)}{\tau^3} & \dots & 0\\
~ & ~ & \ddots & ~\\
0 & 0 & \dots & p(\t)+\frac{2}{d-2}\frac{\eta(\tau)}{\tau}
\end{array}\right) \ 
\ee
Choosing $\nu=\t$ in the conservation equation of (\ref{Stensor}), we obtain
\be\label{Scon}
\p_\t \varepsilon +\frac{d-1}{d-2}\frac{\varepsilon}{\tau} - 2\frac{d-3}{d-2}\frac{\eta}{\tau^2}  =0
\ee
Assuming the viscosity to be a subleading effect, we deduce from (\ref{Scon}) the leading behavior
of the energy density
\be\label{eqbje} \varepsilon \approx \frac{\varepsilon_0}{ \tau^{(d-1)/(d-2)}} \ee
and hence of the pressure, temperature (from the Stefan-Boltzmann law in $d-1$ dimensions, $\varepsilon \sim T^{d-1}$) and entropy density, respectively,
\be\label{eqbjp} p = \frac{\varepsilon}{d-2} \approx \frac{\varepsilon_0}{d-2}\ \frac{1}{\tau^{(d-1)/(d-2)}} \ \ , \ \ \ \
T \approx \frac{T_0}{ \tau^{1/(d-2)}} \ \ , \ \ \ \ s = \frac{dp}{dT} \approx \frac{s_0}{\tau} \ \ , \ \ \ \ s_0 = \frac{d-1}{d-2}\, \frac{\varepsilon_0}{T_0} \ee
The constants $\varepsilon_0$ and $T_0$ represent the initial values of the energy density and temperature, respectively (at $\tau = 1$).

At high temperatures we expect the viscosity to have the same dependence on the temperature as the entropy density (which is known to be true in the case of sinusoidal perturbations of the static $\mathcal{N} = 4$ SYM plasma in five dimensions \cite{son}) so that the ratio $\eta /s$ asymptotes to a constant. Therefore, we shall assume
\be\label{HydroVis}
\eta(\tau)\approx \frac{\eta_0}{\tau}
\ee
where $\eta_0$ is a constant.  We may then solve eq.~(\ref{Scon}) and obtain a subleading correction to the energy density,
\be\label{HydroEP}
\varepsilon = \frac{\varepsilon_0}{\tau^{\frac{d-1}{d-2}}}-\frac{2\eta_0}{\tau^2} + \dots
\ee
yielding corresponding corrections to the temperature and entropy density,
\be\label{HydroT}
T= T_0\left(\frac{1}{\tau^{1/(d-2)}}-\frac{2\eta_0}{(d-1)\varepsilon_0 \tau} + \dots \right)
\ \ , \ \ \ \ s =\frac{dp}{dT}= s_0\left(\frac{1}{\tau}-\frac{2(d-2)\eta_0}{(d-1)\varepsilon_0}\frac{1}{\tau^{\frac{2d-5}{d-2}}} + \dots \right)\ee
Note that to leading order we obtain the ratio
\be
\frac{\eta}{s} = \frac{\eta_0}{s_0} = \frac{(d-2)\eta_0 T_0}{(d-1)\varepsilon_0}
\ee
This ratio is known to take the value $1/(4\pi)$ (eq.~(\ref{eq1})) if the gauge theory fluid is dual to a perturbed AdS Schwarzschild black hole \cite{son}.
We shall show that the above Bjorken fluid also admits a gravity dual which is an appropriately perturbed AdS Schwarzschild black hole leading to the same value (\ref{eq1}) of the viscosity to entropy density ratio.

\section{AdS Schwarzschild black hole}
\label{sec3}

We are interested in obtaining the Bjorken flow described in section \ref{sec2}
as a dual to a large AdS Schwarzschild black hole.
The latter
is a solution of the Einstein equations
\be\label{eqEin}
R_\mu^{~\nu}-\left(\frac{1}{2}R-\Lambda_{d}\right)\delta_\mu^{~\nu}=0
\ee
where $\Lambda_{d} = -\frac{(d-1)(d-2)}{2}$ is a negative cosmological constant.
The solution best suited for our purposes is a large black hole with a flat horizon. Its metric can be written in the form
\be\label{eqmebh} ds_{\mathrm{b.h.}}^2 = \frac{1}{z^2} \left[ -\left( 1 - 2\mu z^{d-1} \right) dt^2 + {d\vec x\,}^2 + \frac{dz^2}{ 1 - 2\mu z^{d-1}} \right] \ee
where $\vec x \in \mathbb{R}^{d-2}$ and $\mu$ is an arbitrary integration constant.
The Hawking temperature of the hole is given by
\be\label{eqTH} T_H = \frac{d-1}{4\pi z_+} \ee
where $z_+$ is the horizon located at
\be\label{eq4} z_+ = (2\mu)^{-\frac{1}{d-1}} \ee
and the boundary of the asymptotically AdS space is at $z=0$.

The VEV of the stress-energy tensor of the dual gauge theory can be constructed using holographic renormalization \cite{Skenderis}.  To this end, the metric needs to be brought in the form of a generally asymptotic AdS metric in Fefferman-Graham coordinates
\be\label{FGa}
ds^2=\frac{g_{\mu\nu}dx^\mu dx^\nu+dz_{FG}^2}{z_{FG}^2}\ee
Then near the
boundary at $z_{FG}=0$, $g_{\mu\nu}$ may be expanded as
\be
g_{\mu\nu} = g_{\mu\nu}^{(0)}+z_{FG}^2 g_{\mu\nu}^{(2)} +
\dots+z_{FG}^{d-1}g_{\mu\nu}^{(d-1)} +\mathcal{O}(z_{FG}^d) \ee
where $g^{(0)}_{\mu\nu}=\eta_{\mu\nu}$.  For the black hole metric (\ref{eqmebh})
the transformation to Fefferman-Graham coordinates is achieved by writing the radial distance in the bulk as
\be\label{eq10a} z = \frac{z_{FG}}{ \left( 1 + \frac{\mu}{2} z_{FG}^{d-1} \right)^{2/(d-1)}} \ee
This brings the black hole metric in the form (\ref{FGa}) with a diagonal $g_{\mu\nu}$ where
\be g_{tt} = - \frac{(1- \frac{\mu}{2} z_{FG}^{d-1})^2}{(1+\frac{\mu}{2} z_{FG}^{d-1})^{2(d-3)/(d-1)}} \ \ , \ \ \ \ g_{ii} = \left( 1 + \frac{\mu}{2} z_{FG}^{d-1} \right)^{4/(d-1)} \ \ \ \ \ (i = 1,\dots, d-2) \ee
The dual gauge theory fluid is static and lives on the $(d-1)$-dimensional Minkowski space
\be\label{eq3} ds_{\mathrm{static}}^2 = g_{\mu\nu}^{(0)} dx^\mu dx^\nu = \eta_{\mu\nu} dx^\mu dx^\nu = - dt^2 + {d\vec x\, }^2 \ee
Evidently, the first correction to the boundary metric is $g_{\mu\nu}^{(d-1)}$,
leading to the stress-energy tensor of the fluid \cite{Skenderis}
\be\label{eqgTa} \langle T_{\mu\nu}\rangle = \frac{d-1}{16\pi G_d} g_{\mu\nu}^{(d-1)}
\ee
where $G_d$ is Newton's constant in the bulk.
Explicitly, we obtain the energy density and pressure, respectively,
\be\label{EPa} \varepsilon = \langle T^{tt} \rangle = \varepsilon_0 \ \ , \ \ \ \
 p = \langle T^{ii} \rangle = \frac{\varepsilon_0}{d-2} \ \ , \ \ \ \ \varepsilon_0 = \frac{(d-2)\mu}{8\pi G_d}\ee
obeying $p = \frac{1}{d-2} \varepsilon$, as expected for a conformal fluid (eq.~(\ref{eqpz})).
This is a perfect fluid and both coefficients of viscosity vanish ($\eta = \zeta = 0$).
With the temperature of the gauge theory fluid coinciding with the Hawking temperature (\ref{eqTH}), we deduce the equation of state
\be p = \frac{1}{16\pi G_d} \left( \frac{4\pi T_H}{d-1} \right)^{d-1} \ee
and the energy and entropy densities, respectively, as functions of temperature,
\be \varepsilon = \frac{d-2}{16\pi G_d} \left( \frac{4\pi T_H}{d-1} \right)^{d-1} \ \ , \ \ \ \ s = \frac{dp}{dT_H} = \frac{1}{4G_d} \left( \frac{4\pi T_H}{d-1} \right)^{d-2} \ee
Since we are interested in Bjorken flow rather than a static fluid, we need to replace the transformation to Fefferman-Graham coordinates (\ref{eq10a}) with one that will yield a metric on the boundary ($g_{\mu\nu}^{(0)}$) of the form (\ref{eq2}) rather than (\ref{eq3}). This is the task we turn to next.

\subsection{Leading order}

Near the boundary, the black hole metric (\ref{eqmebh}) approaches
\be\label{eq5}
ds_{\mathrm{b.h.}}^2 \to \frac{1}{z^2} \left( ds_{\mathrm{static}}^2+dz^2 \right)
\ee
leading to the boundary metric (\ref{eq3}).
For Bjorken flow, we want instead,
\be\label{eq5bj}
ds_{\mathrm{b.h.}}^2 \to \frac{1}{\tilde z^2} \left( ds_{\mathrm{Bjorken}}^2+d\tilde z^2 \right)
\ee
leading to the boundary metric (\ref{eq2}).
We therefore need to find a transformation mapping the black hole coordinates $(t, \vec x, z)$ to a new coordinate system $(\tau, y, \tilde x^i, \tilde z)$ in a patch which includes the boundary relating the two asymptotic forms (\ref{eq5}) and (\ref{eq5bj}). This will be done in the large $\tau$ limit keeping the ratio
\be\label{eqv} v = \frac{\tilde z}{\tau^{1/(d-2)}} \ee
fixed \cite{JP,AS}. More precisely, we define the large $\tau$ limit as follows.
Let
\be \tau = \tau_0 + \tau' \ee
with constant $\tau_0 \gg 1$ and $\tau' \sim \mathcal{O} (1)$, so that $d\tau = d\tau' \sim \mathcal{O} (1)$.
For the transverse dimensions, let $\tilde x^\bot \sim \mathcal{O} (1)$ so that
$(d\tilde x^\perp)^2 \sim \mathcal{O} (1)$.
The remaining term in the boundary metric (\ref{eq2}) will be $\mathcal{O} (1)$ provided we choose the rapidity $y\sim \mathcal{O} (1/\tau)$.
Finally, for the bulk dimension let 
\be \tilde z = \tilde z_0 \tau_0^{1/(d-2)} + \tilde z' \ee
with $\tilde z_0, \tilde z' \sim \mathcal{O} (1)$ to ensure $d\tilde z = d\tilde z' \sim \mathcal{O} (1)$ and $v \sim \mathcal{O} (1)$.

If we now apply the transformation \cite{AS}
\be\label{eq6}
t=\frac{d-2}{d-3} \tau^{\frac{d-3}{d-2}}~~,~~~~x^1=\tau^{\frac{d-3}{d-2}} y~~,~~~~
x^\bot=\frac{\tilde x^\bot}{\tau^{1/(d-2)}}~~,~~~~z=\frac{\tilde z}{\tau^{1/(d-2)}}
\ee
where $x^\perp = (x^2,\dots,x^{d-2})$,
to the black hole metric (\ref{eqmebh}) (more presicely to a patch which includes the boundary $z\to 0$), we arrive at
\be\label{TransMetric}
ds^2_{\mathrm{b.h.}} = \frac{1}{\tilde z^2} \left[ - \left( 1-2\mu  v^{d-1} \right) d\tau^2 + \tau^2 dy^2 + (d\tilde x^\bot)^2 + \frac{d\tilde z^2}{1-2\mu v^{d-1}} \right] + \mathcal{O} (1/\tau^{(d-1)/(d-2)})
\ee
with $v$ defined in (\ref{eqv}).
The new coordinate system defines a new foliation near the boundary of $\tilde z=$~const.~hypersurfaces with time-dependent metrics leading to a flow on the boundary. At leading order in $\tau$, near the boundary the black hole metric behaves as in (\ref{eq5bj}), as desired.

To see the form of the flow of the gauge theory fluid on the boundary, we need to bring (\ref{TransMetric}) in Fefferman-Graham form. This can be done to leading order in the large $\tau$ expansion via the change of coordinates
\be\label{eq10} \tilde z = z_{FG} \left[ 1 - \frac{\mu}{d-1} \frac{z_{FG}^{d-1}}{\tau^{(d-1)/(d-2)}} + \mathcal{O} (z_{FG}^{2(d-1)}) \right] \ee
This brings the metric (\ref{TransMetric}) in the form (\ref{FGa}) with a diagonal $g_{\mu\nu}$ whose value at the boundary ($g_{\mu\nu}^{(0)}$) is given by (\ref{eq2}). The first correction has non-vanishing components
\be\label{eqgd} g_{\tau\tau}^{(d-1)} = 2\mu \frac{d-2}{d-1} \ \frac{1}{\tau^{(d-1)/(d-2)}}
\ \ , \ \ \ \ g_{ii}^{(d-1)} = \frac{1}{\tau^2} g_{yy}^{(d-1)} = \frac{2\mu}{d-1}\ \frac{1}{\tau^{(d-1)/(d-2)}} \ee
This metric agrees with the asymptotic time-dependent solution of the Einstein equations obtained by Janik and Peschanski \cite{JP} in five dimensions.
It leads to a stress-energy tensor for the conformal fluid on the boundary via holographic renormalization (eq.~(\ref{eqgTa})) in agreement with expectations from Bjorken hydrodynamics (eqs.~(\ref{eqbje}) and (\ref{eqbjp})), where
\be\label{EP} \varepsilon_0 = \frac{(d-2)\mu}{8\pi G_d} \ee
Notice also that the resulting energy and pressure agree with the case of a static boundary (\ref{EPa}) at initial time $\tau = 1$.

The temperature of the gauge theory fluid can also be determined as it is related to the Hawking temperature (\ref{eqTH}) of the black hole.
To see this, note that the static metric (\ref{eq3}) is conformally equivalent to the Bjorken metric (\ref{eq2}). Indeed, applying the restriction of the transformation (\ref{eq6}) on the boundary,
\be\label{eq6a}
t=\frac{d-2}{d-3} \tau^{\frac{d-3}{d-2}}~~,~~~~x^1=\tau^{\frac{d-3}{d-2}} y~~,~~~~
x^\bot=\frac{\tilde x^\bot}{\tau^{1/(d-2)}}
\ee
we obtain at leading order in $\tau$,
\be
ds_{\mathrm{static}}^2 = \tau^{-\frac{2}{d-2}}\left[ ds_{\mathrm{Bjorken}}^2 + \mathcal{O}(1/\tau)\right]
\ee
The conformal factor, $\tau^{-1/(d-2)}$, scales the inverse of the Euclidean proper time period of the thermal Green function on the Bjorken boundary and hence the temperature of the gauge theory fluid, i.e.,
\be\label{eqTHLO}
T = \frac{T_H}{\tau^{1/(d-2)}}
\ee
where the proportionality constant is chosen so that the temperature agrees with the temperature of the static fluid ($T_H$) initially (at $\tau =1$).
This is in agreement with the hydrodynamic result (\ref{eqbjp}) with
\be T_0 = T_H \ee
Finally, the entropy also agrees with (\ref{eqbjp}) where
\be s_0 = \frac{(2\mu)^{(d-2)/(d-1)}}{4G_d} \ee
The leading-order metric (\ref{TransMetric}) can be seen to be regular in the bulk. Indeed, the Kretschmann scalar
\be\label{eqKS} \mathcal{R}^2 = R_{ABCD} R^{ABCD} \ee
is found to be
\be\label{eqKS0} \mathcal{R}^2 = 2(d-1)\left[ d+2(d-2)^2(d-3)\mu^2 v^{2(d-1)} \right] + \mathcal{O} (1/\tau^{(d-3)/(d-2)} ) \ee
whose only singularity is obtained in the limit $z\to\infty$.

Moreover, one can construct invariants which are linear combinations of the components of the Riemann tensor. For the geometry to be regular, these invariants must also be free of singularities \cite{BBB1}.
Introducing the vielbein
\be e_a^A = \left\{ \left( \begin{array}{c} \tilde z/\sqrt{1-2\mu v^{d-1}} \\ 0 \\ 0 \\ 0 \\ 0 \end{array} \right) \ , \ \
 \left( \begin{array}{c} 0 \\ \tilde z/\tau \\ 0 \\ 0 \\ 0 \end{array} \right) \ , \ \
 \left( \begin{array}{c} 0 \\ 0 \\ \tilde z \\ 0 \\ 0 \end{array} \right) \ , \ \
 \dots  \ , \ \
 \left( \begin{array}{c} 0 \\ 0 \\ 0 \\ 0 \\ \sqrt{1-2\mu v^{d-1}} /\tilde z \end{array} \right)
\right\} \ee
with $a=0,\dots,d-1$ spanning a local Minkowski space such that $g_{AB}e^A_a e^B_b=\eta_{ab}$.  The Riemann tensor invariants are
\be\label{eq42} \mathcal{R}_{abcd} = R_{ABCD} e_a^A e_b^B e_c^C e_d^D \ee
Even though the individual components of the Riemann tensor have singularities,
the invariants (\ref{eq42}) are all regular. The non-vanishing components are found to be
\bes
\mathcal{R}_{1 0 1 0} = \mathcal{R}_{2 0 2 0} =\dots = \mathcal{R}_{(d-2) 0 (d-2) 0} = 1+(d-3)\mu v^{d-1}+ \mathcal{O} (1/\tau^{(d-3)/(d-2)})\nonumber\\
\mathcal{R}_{1 2 1 2} = \dots=\mathcal{R}_{(d-2) 1 (d-2) 1} = \dots=\mathcal{R}_{(d-3) (d-2) (d-3) (d-2)} = -1 + 2\mu v^{d-1}+ \mathcal{O} (1/\tau^{(d-3)/(d-2)})\nonumber\\
\mathcal{R}_{(d-1) 1 (d-1) 1} = \dots = \mathcal{R}_{(d-1) (d-2) (d-1) (d-2)}  = -1 - (d-3)\mu v^{d-1}+ \mathcal{O} (1/\tau^{(d-3)/(d-2)})\nonumber\\
\mathcal{R}_{(d-1) 0 (d-1) 0} = 1-(d-3)(d-2)\mu v^{d-1}+ \mathcal{O} (1/\tau^{(d-3)/(d-2)})
\ees
together with those obtained using the symmetries of the Riemann tensor.

\subsection{Next-to-leading order}

To extend the above results to next-to-leading order ($\mathcal{O} (1/\tau^{(d-3)/(d-2)})$), let us add a correction to the transformation (\ref{eq6}) so that it reads
\bes\label{eq34}
t&=&\tau^{\frac{d-3}{d-2}}\left(\frac{d-2}{d-3}+\frac{(d-3)\tau^2 y^2-(\tilde x^\bot)^2}{2(d-2)\tau^2(1-2\mu v^{d-1})}\right)-\mathcal{C}_1\ln\tau + \frac{f_1(v)}{\tau^{(d-3)/(d-2)}}\nonumber\\
x^1&=&\tau y \left( \frac{1}{\tau^{1/(d-2)}} -\frac{\mathcal{C}_1+b_1(v)}{\tau}\right)~~,~~~~x^\bot=\tilde x^\perp \left( \frac{1}{\tau^{1/(d-2)}} -\frac{\mathcal{C}_1+c_1(v)}{\tau}\right)\nonumber\\
z&=&\tilde z  \left( \frac{1}{\tau^{1/(d-2)}}-\frac{\mathcal{C}_1}{\tau}\right)
\ees
where $\mathcal{C}_1$ is an arbitrary constant, $v$ is defined in (\ref{eqv}) and $b_1(v)$, $c_1(v)$ and $f_1(v)$ are functions which vanish at the boundary ($v=0$), so that they do not alter the boundary behavior of the metric obtained at leading order above.

The function $f_1(v)$ is determined by the requirement that the $\tau\tilde z$ component of the metric vanish at the order we are interested in. We obtain the constraint
\be
v+(d-2)\left(1-2\mu v^{d-1}\right)^2 f_1'(v)=0
\ee
whose unique solution (with $f_1(0)=0$)
may be written in terms of a hypergeometric function,
\be\label{eqf1}
f_1(v)=\frac{-v^2(d-3)}{2(d-2)(d-1)} F \left(1,\frac{2}{d-1};\frac{d+1}{d-1};2\mu v^{d-1}\right)-\frac{v^2}{(d-2)(d-1)(1-2\mu v^{d-1})}
\ee
With this choice of $f_1(v)$,
under the transformation (\ref{eq34}) the black hole metric (\ref{eqmebh}) turns into
\bes\label{eq45} ds_{\mathrm{b.h.}}^2 &=& \frac{1}{\tilde z^2} \Bigg[ - \left( 1-2\mu v^{d-1} + \frac{2(d-1)\mu\mathcal{C}_1 v^{d-1}}{\tau^{(d-3)/(d-2)}} \right) d\tau^2 + \left( 1 -\frac{2b_1(v)}{\tau^{(d-3)/(d-2)}} \right) \tau^2 dy^2 + \left( 1 -\frac{2c_1(v)}{\tau^{(d-3)/(d-2)}} \right) (d\tilde x^\perp)^2 \nonumber\\
& & + \frac{d\tilde z^2}{ 1-2\mu v^{d-1} + \frac{2(d-1)\mu\mathcal{C}_1 v^{d-1}}{\tau^{(d-3)/(d-2)}}} +\dots \Bigg] \ees
where the dots represent higher-order terms in the large $\tau$ expansion.

The Einstein equations at next-to-leading order yield two independent equations for the functions $b_1(v)$, $c_1(v)$,
\bes
b_1'(v)+(d-3)c_1'(v)&=& 0\nonumber\\
-\mu\mathcal{C}_1 v^{d-2} + \frac{\mu v^{d-2}}{d-2}\left[ b_1(v)+(d-3)c_1(v)\right] +\frac{1-2\mu v^{d-1}}{(d-1)(d-3)} b_1'(v) &=& 0
\ees
whose unique solution (with the boundary conditions $b_1(0)=c_1(0)=0$) is
%
\be\label{eq47}
b_1(v)=-\frac{(d-3)\mathcal{C}_1}{2}\ln\left(1-2\mu v^{d-1}\right)~~,~~~~c_1(v)=\frac{\mathcal{C}_1}{2}\ln\left(1-2\mu v^{d-1}\right)
\ee
Using (\ref{eq47}), the metric (\ref{eq45}) can be written as
\bes\label{eq48} ds_{\mathrm{b.h.}}^2 &=& \frac{1}{\tilde z^2} \Bigg[ - \left( 1-2\mu v^{d-1} + \frac{2(d-1)\mu\mathcal{C}_1 v^{d-1}}{\tau^{(d-3)/(d-2)}} \right) d\tau^2 + \left( 1 -2\mu v^{d-1} \right)^{(d-3)\mathcal{C}_1/\tau^{(d-3)/(d-2)}} \tau^2 dy^2 \nonumber\\
& & + \left( 1 -2\mu v^{d-1} \right)^{-\mathcal{C}_1/\tau^{(d-3)/(d-2)}} (d\tilde x^\perp)^2 + \frac{d\tilde z^2}{ 1-2\mu v^{d-1} + \frac{2(d-1)\mu\mathcal{C}_1 v^{d-1}}{\tau^{(d-3)/(d-2)}}} +\dots \Bigg] \ees
which includes $\mathcal{O} (1/\tau^{(d-3)/(d-2)})$ corrections to the leading-order expression (\ref{TransMetric}). The next-to-leading-order expression (\ref{eq48}) for the metric has no dependence on the rapidity $y$ and transverse coordinates $\tilde x^\perp$, therefore it leads to a Bjorken flow for the gauge theory fluid on the boundary.

We may now use holographic renormalization \cite{Skenderis} to calculate the VEV of the stress-energy tensor of the dual gauge theory.
The transformation to Fefferman-Graham coordinates (\ref{FGa}) is
\be\label{eq30} \tilde z = z_{FG} \left[ 1 - \mu \left( \frac{1}{d-1} - \frac{\mathcal{C}_1}{\tau^{(d-3)/(d-2)}} \right) \frac{z_{FG}^{d-1}}{\tau^{\frac{d-1}{d-2}}}  + \mathcal{O} (z_{FG}^{2(d-1)}) \right] \ee
correcting the leading-order transformation (\ref{eq10}).
The form of the boundary metric is unaltered by design whereas the first non-vanishing correction away from the boundary reads
\bes\label{eqgdNLO} g_{\tau\tau}^{(d-1)} &=& \frac{2\mu(d-2)}{d-1} \left(\frac{1}{\tau^{(d-1)/(d-2)}} - \frac{(d-1)\mathcal{C}_1}{\tau^{2}} \right) \nonumber\\ 
g_{ii}^{(d-1)} = \frac{2\mu}{d-1}\frac{1}{\tau^{(d-1)/(d-2)}}\ &,& \ \ \ \frac{1}{\tau^2} g_{yy}^{(d-1)} = \frac{2\mu}{d-1}\left(\frac{1}{\tau^{(d-1)/(d-2)}}-\frac{(d-1)(d-2)\mathcal{C}_1}{\tau^2}\right) \ees
correcting the leading-order expression (\ref{eqgd}).

Using eq.~(\ref{eqgTa}), we obtain a stress-energy tensor for the conformal fluid in agreement with Bjorken hydrodynamics (eqs.~(\ref{HydroVis}) and (\ref{HydroEP})) with $\varepsilon_0$ as before (eq.~(\ref{EP})) and
\be\label{eqetaC} \eta_0 = \frac{(d-1)\mathcal{C}_1\varepsilon_0}{2} \ee
matching the result of ref.~\cite{NakSin} for $d=5$.

The temperature of the gauge theory fluid can also be determined through the conformal factor relating the static metric (\ref{eq3}) to the Bjorken metric (\ref{eq2}), as before. Applying the restriction of the transformation (\ref{eq34}) on the boundary,
\be\label{eq6aNLO}
t=\frac{d-2}{d-3} \tau^{\frac{d-3}{d-2}} - \mathcal{C}_1 \ln \tau ~~,~~~~x^1=\tau y \left( \frac{1}{\tau^{1/(d-2)}} - \frac{\mathcal{C}_1}{\tau} \right) ~~,~~~~
x^\bot=\tilde x^\bot \left( \frac{1}{\tau^{1/(d-2)}} - \frac{\mathcal{C}_1}{\tau} \right)
\ee
we obtain at next-to-leading order in $\tau$,
\be\label{eqstbjNLO}
ds_{\mathrm{static}}^2 = \left( \frac{1}{\tau^{1/(d-2)}} - \frac{\mathcal{C}_1}{\tau} \right)^2 \left[ ds_{\mathrm{Bjorken}}^2 + \dots \right]
\ee
which yields the $\tau$-dependent temperature
\be\label{eqTNLO}
T = T_H \left( \frac{1}{\tau^{1/(d-2)}} - \frac{\mathcal{C}_1}{\tau} \right)
\ee
correcting the leading-order result (\ref{eqTHLO}) and in agreement with the hydrodynamic result (\ref{HydroT}) with $T_0 = T_H$.
The correct expression for the entropy density also follows and we obtain the ratio
\be\label{eqetaNLO} \frac{\eta}{s} = \frac{(d-1)(d-2)}{8\pi}\ \mathcal{C}_1 (2\mu)^{1/(d-1)} \ee
There is no constraint on this ratio at this order because the truncated metric (\ref{eq48}) is regular in the bulk \cite{NakSin}.
This can be seen by a calculation of the Kretschmann scalar (\ref{eqKS}).
With the metric (\ref{eq48}), we obtain
\be\label{eqKS1} \mathcal{R}^2 = 2(d-1)\left[ d+2(d-2)^2(d-3)\mu^2 v^{2(d-1)}\left(1-\frac{2(d-1)\mathcal{C}_1}{\tau^{(d-3)/(d-2)}}\right) \right] + \dots \ee
Eq.~(\ref{eqKS1}) corrects the leading-order result (\ref{eqKS0}) showing that to this order the Kretschmann scalar is regular.

However, the metric (\ref{eq48}) leads to singular Riemann invariants (\ref{eq42}). Indeed, we obtain explicitly, e.g., for $d=5$,
\be\label{eq62} \mathcal{R}_{0101} = 1 + 2\mu v^{4} + \frac{32\mathcal{C}_1 \mu^2 v^{8}}{\tau^{2/3}}\ \frac{1}{1-2\mu v^{4}} + \dots \ee
exhibiting a simple pole at $v = (2\mu)^{-1/4}$.
This singularity should be absent, since our metric comes from a Schwarzschild black hole which has no singularities except as $z\to\infty$. We obtained a pole because we have not included all contributions at order $\mathcal{O} (1/\tau^{(d-3)/(d-2)})$. There are additional contributions from next order ($\mathcal{O} (1/\tau)$) terms are in the metric (\ref{eq48}).
Including them, the corrected metric reads
\bes\label{eq48c} ds_{\mathrm{b.h.}}^2 &=& \frac{1}{\tilde z^2} \Bigg[ - \left( 1-2\mu v^{d-1} + \frac{2(d-1)\mu\mathcal{C}_1 v^{d-1}}{\tau^{(d-3)/(d-2)}} \right) d\tau^2 + \left( 1 -2\mu v^{d-1} \right)^{(d-3)\mathcal{C}_1/\tau^{(d-3)/(d-2)}} \tau^2 dy^2 \nonumber\\
& & + \left( 1 -2\mu v^{d-1} \right)^{-\mathcal{C}_1/\tau^{(d-3)/(d-2)}} (d\tilde x^\perp)^2 + \frac{d\tilde z^2}{ 1-2\mu v^{d-1} + \frac{2(d-1)\mu\mathcal{C}_1 v^{d-1}}{\tau^{(d-3)/(d-2)}}} +2\mathcal{A}_\mu d\tilde x^\mu d\tilde z+2\mathcal{B}_\mu d\tilde x^\mu d\tau + \dots \Bigg] \ees
where the off-diagonal elements are
\be \mathcal{A}_\tau = 0 \ , \ \ \mathcal{A}_y = -\frac{(d-1)(d-3)\mathcal{C}_1\mu \tau y v^{d-2}}{1-2\mu v^{d-1}} \ , \ \ \mathcal{A}_{\tilde x^\perp} = \frac{(d-1)\mathcal{C}_1 \mu \tilde x^\perp v^{d-2}}{\tau (1-2\mu v^{d-1})} \ee
These corrections do not lead to a Bjorken flow. However, the metric (\ref{eq48c}) satisfies the Einstein equations at this order. The Kretschmann scalar (\ref{eqKS1}) is unaltered, and the Riemann invariants (\ref{eq42}) are corrected with the corrections cancelling all singularities. E.g., the invariant (\ref{eq62}) for $d=5$ is corrected to
\be\label{eq62a} \mathcal{R}_{0101} = 1 + 2\mu v^{4} - \frac{8\mathcal{C}_1 \mu v^{4}}{\tau^{2/3}} + \dots \ee
which is a regular expression.


\subsection{Next-to-next-to-leading order}

Extending the above results to next-to-next-to-leading order requires calculations which are considerably involved.
We shall therefore restrict attention to the physically interesting case of five dimensions, setting $d=5$, and employ {\sf Mathematica} for the lengthy algebraic manipulations. The generalization to an arbitrary dimension is straightforward but adds little to the main results.

Let us augment the transformation (\ref{eq34}) for $d=5$ with appropriate $\mathcal{O} (1/\tau^{4/3})$ terms as follows,
\bes\label{eq34a}
t&=&\frac{3}{2} \tau^{2/3} \left[ 1 + \frac{2\tau^2 y^2 - (\tilde x^\perp)^2}{9(1-2\mu v^4) \tau^2} \right] -\mathcal{C}_1\ln\tau + \frac{f_1(v)-\frac{3}{2}\mathcal{C}_2}{\tau^{2/3}} + \frac{f_2(v)}{\tau^{4/3}} \nonumber  \\
x^1&=&\tau^{2/3} y \left( 1 -\frac{\mathcal{C}_1+b_1(v)}{\tau^{2/3}} + \frac{b_2(v)+\mathcal{C}_2}{\tau^{4/3}}  \right)\nonumber\\
x^\bot&=& \frac{\tilde x^\perp}{\tau^{1/3}} \left( 1-\frac{\mathcal{C}_1+c_1(v)}{\tau^{2/3}} + \frac{c_2(v)+\mathcal{C}_2}{\tau^{4/3}} \right)\nonumber\\
z&=& v  \left( 1-\frac{\mathcal{C}_1}{\tau^{2/3}} + \frac{a_2(v)+\mathcal{C}_2}{\tau^{4/3}} \right)
\ees
where $v$ is defined in (\ref{eqv}).  The constant $\mathcal{C}_1$ is once again related to the viscosity coefficient (eq.~(\ref{eqetaC})), but to understand $\mathcal{C}_2$ one must employ second order hydrodynamics \cite{Hydro2}. Whereby it is understood to be related to the relaxation time.
The functions $f_1(v)$ and $b_1(v), c_1(v)$ have already been determined at first perturbative order (eqs.~(\ref{eqf1}) and (\ref{eq47}), respectively, with $d=5$).
The new functions $f_2 (v)$ and $a_2(v), b_2(v), c_2(v)$ ought to vanish at the boundary ($v=0$) so as not to contribute to the boundary metric.

As with $f_1(v)$, demanding that the $\tau\tilde z$ component of the metric vanish at the order we are interested in yields the constraint on $f_2(v)$,
\be 3(1-2\mu v^4)^3 f_2'(v) -\mathcal{C}_1 v (3 +10\mu v^4 ) = 0 \ee
which has the unique solution (with $f_2(0)=0$)
\be f_2(v) = -\mathcal{C}_1 f_1(v) + \frac{\mathcal{C}_1 v^2}{3(1-2\mu v^4)^2} \ee
With this choice of $f_2(v)$, the application of the transformation (\ref{eq34a}) to the black hole metric (\ref{eqmebh}) with $d=5$ turns the latter into the form
\bes\label{eqbhNNLO} ds_{\mathrm{b.h.}}^2 &=& \frac{1}{\tilde z^2} \Bigg[ - \left( 1-2\mu v^{4} + \frac{8\mu\mathcal{C}_1v^{4}}{\tau^{2/3}} + \frac{A_2(v) }{\tau^{4/3}} \right) d\tau^2
+ \left( 1 -\frac{2b_1(v)}{\tau^{2/3}} + \frac{B_2(v)}{\tau^{4/3}} \right) \tau^2 dy^2\nonumber\\
& & + \left( 1 -\frac{2c_1(v)}{\tau^{2/3}} + \frac{C_2(v)}{\tau^{4/3}} \right) (d\tilde x^\perp)^2
+ \frac{d\tilde z^2}{1-2\mu v^{4} + \frac{8\mu\mathcal{C}_1v^{4}}{\tau^{2/3}} - \frac{d_2(v)}{\tau^{4/3}} }
+ 2\mathcal{A}_\mu d\tilde x^\mu d\tilde z +\dots \Bigg] \ees
where the dots represent higher-order terms.

The off-diagonal elements are
\be \mathcal{A}_\tau = \frac{4\mu v^3 ((\tilde x^\perp)^2 -2\tau^2 y^2)}{3(1-2\mu v^4) \tau^{4/3}} \ \ , \ \ \ \ \mathcal{A}_y = - \frac{8\mathcal{C}_1\mu \tau y v^3}{1-2\mu v^4} \ \ , \ \ \ \ \mathcal{A}_{\tilde x^\perp} = \frac{4\mathcal{C}_1\mu \tilde x^\perp v^3}{\tau (1-2\mu v^4)} \ee
and we have defined
\bes A_2(v) &=& \frac{v^2}{9(1-2\mu v^4)} - 4\mu v^4(3\mathcal{C}_1^2 +2\mathcal{C}_2) -\frac{4}{3}(1-2\mu v^4) f_1(v) - 2(1+2\mu v^4) a_2(v) \nonumber\\
B_2(v) &=& b_1^2(v)-2\mathcal{C}_1 b_1(v) - 2a_2(v) + 2b_2(v) \nonumber\\
C_2(v) &=& c_1^2(v) -2\mathcal{C}_1 c_1(v) - 2a_2(v) + 2c_2(v) \nonumber\\
d_2(v) &=& 4\mu v^4 (3\mathcal{C}_1^2 +2\mathcal{C}_2+ 2 a_2(v)) + 2v(1-2\mu v^4)a_2'(v)
- \frac{v^2}{9(1-2\mu v^4)} \ees
Evidently, the metric depends on the rapidity as well as the transverse coordinates at next-to-next-to-leading order. This dependence cannot be eliminated by any choice of the functions which are yet to be determined. One may try to modify the transformation (\ref{eq34a}) to eliminate the off-diagonal terms which depend on $y$ and $\tilde x^\perp$, but this only shifts the dependence on these coordinates to other components of the metric.
If we insist on reproducing Bjorken flow on the boundary, we must perturb the Schwarzschild metric (\ref{eqmebh}). Let the perturbed metric be
\be\label{eqpert} ds_{\mathrm{perturbed}}^2 = ds_{\mathrm{b.h.}}^2 - \frac{1}{\tilde z^2} \left[ \frac{v^2\mathcal{A} (v)}{\tau^{4/3}} d\tilde z^2 + 2\mathcal{A}_\mu d\tilde x^\mu d\tilde z \right] \ee
where, apart from the off-diagonal elements, we are also modifying the $\tilde z\tilde z$ component of the black hole metric by an amount proportional to an arbitrary function $\mathcal{A} (v)$.
It turns out that, even though we have certain freedom in the choice of $\mathcal{A} (v)$ (gauge freedom), this function cannot vanish.

Using eqs.~(\ref{eqbhNNLO}) and (\ref{eqpert}), the $\tilde z\tilde z$ component of the perturbed metric can be expanded as
\be g_{\tilde z\tilde z} = \frac{1}{\tilde z^2} \left[ \frac{1}{1-2\mu v^4}
- \frac{8\mu\mathcal{C}_1 v^{4}}{\tau^{2/3} (1-2\mu v^{4})^2}
+ \frac{v^2 D_2(v)}{\tau^{4/3}} + \dots \right] \ee
where
\be D_2(v) = \frac{d_2(v)}{v^2(1-2\mu v^4)^2} + \frac{64\mu^2 \mathcal{C}_1^2 v^6}{(1-2\mu v^4)^3} - \mathcal{A} (v)\ee
Demanding that the perturbed metric satisfy the Einstein equations at the order we are interested in yields four constraints on the four functions $A_2(v),B_2(v),C_2(v),D_2(v)$,
\bes 3(1-2\mu v^4)^2 (3-2\mu v^4) (B_2'+2C_2') -3v(1-2\mu v^4)^3 (B_2'' +2 C_2'') & & \nonumber\\
-9v^2 (1-2\mu v^4)^4 D_2' + 18v (1+6\mu v^4) (1-2\mu v^4)^3 D_2 & & \nonumber\\
+ 8\mu v^5 \left[ -1 + 36\mathcal{C}_1^2 \mu v^2 \left( 11-6\mu v^4 - 2(1-2\mu v^4) \ln (1-2\mu v^4)\right) \right] & = & 0 \nonumber\\
9v (1-2\mu v^4)^3 A_2'' - 9 (3-10\mu v^4)(1-2\mu v^4)^2 A_2' +288 \mu^2 v^7 (1-2\mu v^4) A_2 & & \nonumber\\
+9v^2 (3-2\mu v^4)(1-2\mu v^4)^4 D_2' -18v (3+8\mu v^4 -12\mu^2 v^8) (1-2\mu v^4)^3 D_2 & & \nonumber\\
+18v (1-2\mu v^4)^4 C_2'' -18 (3+2\mu v^4) (1-2\mu v^4)^3 C_2'
-8\mu v^5 \left[ 7+2\mu v^4 +36 \mathcal{C}_1^2 \mu v^2 (9+ 44\mu v^4 + 4\mu^2 v^8)\right] & = & 0 \nonumber\\
9v (1-2\mu v^4)^3 A_2'' - 9 (3-10\mu v^4)(1-2\mu v^4)^2 A_2' +288 \mu^2 v^7 (1-2\mu v^4) A_2 & & \nonumber\\
+9v (1-2\mu v^4)^4 (B_2''+C_2'') -9 (3+2\mu v^4) (1-2\mu v^4)^3 (B_2'+C_2')
& & \nonumber\\
+9v^2 (3-2\mu v^4)(1-2\mu v^4)^4 D_2' -18v (3+8\mu v^4 -12\mu^2 v^8) (1-2\mu v^4)^3 D_2 & & \nonumber\\
-8\mu v^5 \left[ 1+14\mu v^4 +36 \mathcal{C}_1^2 \mu v^2 (39-28\mu v^4 + 28\mu^2 v^8)\right] & = & 0 \nonumber\\
-9(1-2\mu v^4)^2 A_2' - 72\mu v^3 (1-2\mu v^4) A_2 -3(1-2\mu v^4)^2 (3-2\mu v^4) (B_2'+2C_2') -36 v (1-2\mu v^4)^3 D_2 & & \nonumber\\
+ 8\mu v^3 \left[ v^2 + 18\mathcal{C}_1^2 \left( 14\mu v^4 +4\mu^2 v^8 - (1-2\mu v^4) (3-2\mu v^4)\ln (1-2\mu v^4)\right) \right] & = & 0\ees
coming from the $\tau\tau$, $yy$, $xx$ and $zz$ components of the Einstein equations, respectively.
This system of equations does not completely determine the four functions. Keeping $A_2(v)$ arbitrary (gauge degree of freedom), the other three functions are determined to be
\bes B_2'(v) &=&  \left( \frac{A_2(v)}{1+2\mu v^4} \right)' \nonumber\\
& & + \frac{2\mu v^3}{9(1-4\mu^2 v^8)^2} \Bigg\{ -4v^2(3 +4 \mu v^4 +4 \mu^2 v^8) -8(1-2\mu v^4)(1+\mu v^4 + 2\mu^2 v^8)\frac{1}{\sqrt{2\mu}} \tanh^{-1} v^2 \sqrt{2\mu} \nonumber\\
& &-72\mathcal{C}_1^2(1-24\mu v^4-20\mu^2v^8)
-72 \mathcal{C}_1^2 (5+2\mu v^4 + 8\mu^2 v^8)(1-2\mu v^4)\ln (1-2\mu v^4) \nonumber\\
& & +\mathcal{C}_3(1-2\mu v^4)^2-\mathcal{C}_4(1-2\mu v^4)(3+4\mu^2 v^8) \Bigg\} \nonumber\\
C_2' (v) &=&   \left( \frac{A_2(v)}{1+2\mu v^4} \right)' \nonumber\\
& & + \frac{2\mu v^3}{9(1-4\mu^2 v^8)^2} \Bigg\{ -2v^2(3 -4 \mu v^4 -4 \mu^2 v^8) -2(1-10\mu v^4 + 12\mu^2 v^8+8\mu^3 v^{12})\frac{1}{\sqrt{2\mu}} \tanh^{-1} v^2 \sqrt{2\mu} \nonumber\\
& & -36\mathcal{C}_1^2 (11-6\mu v^4+20\mu^2 v^8+24 \mu^3 v^{12})
-36 \mathcal{C}_1^2 (7-22\mu v^4 + 20\mu^2 v^8 -8\mu^3 v^{12})\ln (1-2\mu v^4) \nonumber\\
& & +\mathcal{C}_3 (1-2\mu v^4)^2 +\mathcal{C}_4(-\frac{3}{2}+9\mu v^4-10\mu^2 v^8 - 4\mu^3 v^{12}) \Bigg\}\nonumber\ees
\bes
D_2(v) &=& -\frac{1}{v(1-4\mu^2 v^8)} A_2'(v) + \frac{4\mu v^2 (1-6\mu v^4)}{(1-4\mu^2 v^8)^2} A_2 (v) \nonumber\\
& & + \frac{\mu v^2 (3-2\mu v^4)}{9 (1-2\mu v^4) (1+2\mu v^4)^2} \left[  \frac{2}{\sqrt{2\mu}} \tanh^{-1} v^2\sqrt{2\mu} + 108 \mathcal{C}_1^2 \ln (1-2\mu v^4) -\frac{\mathcal{C}_3}{2} +\mathcal{C}_4\right] \nonumber\\
& & + \frac{2\mu v^2}{9(1-4\mu^2 v^8)^2 (1-2\mu v^4)} \left[ -v^2 (7+4\mu^2 v^8) + 72\mathcal{C}_1^2 (3-6\mu v^4 +20 \mu^2 v^8+24\mu^3 v^{12}) \right]\ees
The functions $B_2(v)$ and $C_2(v)$ are found by integrating the first two equations, respectively. No arbitrary integration constants are introduced because $B_2(0)=C_2(0)=0$. Notice also that apart from the arbitrary function $A_2(v)$, the above functions contain an arbitrary parameters $\mathcal{C}_3$ and $\mathcal{C}_4$.

Constraints on the parameters are obtained by demanding regularity of the perturbed metric in the bulk.
After some algebra, the Kretschmann scalar (\ref{eqKS})
is found as an asymptotic expansion in $\tau$,
\bes \mathcal{R}^2 &=& 8(5+36\mu^2 v^8) - \frac{2304\mathcal{C}_1 \mu^2 v^8}{\tau^{2/3}} \nonumber\\
& & + \frac{96\mu^2 v^8}{9(1+2\mu v^4)\tau^{4/3}} \Bigg\{ -108 A_2 +\frac{v^2 (-14 + 8\mu v^4 - 24\mu^2 v^8) + 72 \mathcal{C}_1^2 (3+24\mu v^4-44\mu^2 v^8+32\mu^3 v^{12})}{(1-2\mu v^4)^2} \nonumber\\
& &(\frac{3}{2}-3\mu v^4)\mathcal{C}_3-3(1-2\mu v^4)\mathcal{C}_4 -6(1-2\mu v^4) \left[ \frac{1}{\sqrt{2\mu}} \tanh^{-1} v^2\sqrt{2\mu} +54 \mathcal{C}_1^2 \ln (1-2\mu v^4) \right] \Bigg\}
 + \dots \ees
correcting the lower order expression (\ref{eqKS1}) for $d=5$.
At this order, we have a
double pole at $v = 1/(2\mu)^{1/4}$. Demanding regularity of the Kretschmann scalar we obtain two constraints. One fixes the parameter $\mathcal{C}_1$ (which is related to the viscosity coefficient),
\be\label{eqCco} \mathcal{C}_1 = \frac{1}{6(2\mu)^{1/4}} \ee
and the other fixes the residue of the function $A_2(v)$ (which ought to have a simple pole at $v= 1/(2\mu)^{1/4}$). Near the pole, we obtain
\be A_2(v) \approx \frac{v^2}{9(1-2\mu v^4)} \ee
Thus, $A_2(v)$ cannot vanish, however, other than the simple pole at $v= 1/(2\mu)^{1/4}$, it is arbitrary.
Finally, there are no constraints on the parameters $\mathcal{C}_3$ and $\mathcal{C}_4$.

The Riemann invariants (\ref{eq42}) are singular, as in lower order (eq.~(\ref{eq62})). Their singularities are canceled by higher-order contributions to the metric (\ref{eqpert}) which, however, spoil boost invariance ({\em cf.}~with the corrected expression (\ref{eq62a}) due to the corrected metric (\ref{eq48c})).

Having obtained an explicit expression for the metric, we may now use holographic renormalization to compute the stress-energy tensor of the dual gauge theory.
This is a tedious task. However, note that the temperature is easy to deduce from the restriction of the transformation (\ref{eq34a}) to the boundary,
\bes\label{eq34ab}
t&=&\frac{3}{2} \tau^{2/3} \left[ 1 + \frac{2\tau^2 y^2 - (\tilde x^\perp)^2}{9 \tau^2} \right] -\mathcal{C}_1\ln\tau -\frac{3\mathcal{C}_2}{2\tau^{2/3}} \nonumber\\
x^1&=&\tau^{2/3} y \left( 1 -\frac{\mathcal{C}_1}{\tau^{2/3}} + \frac{\mathcal{C}_2}{\tau^{4/3}} \right)\ , \ \ 
x^\bot= \frac{\tilde x^\perp}{\tau^{1/3}} \left( 1-\frac{\mathcal{C}_1}{\tau^{2/3}} + \frac{\mathcal{C}_2}{\tau^{4/3}} \right)
\ees
with a new correction dependent on $\mathcal{C}_2$.
The metric perturbation does not change the argument we employed at lower orders because the perturbation vanishes at the boundary.
From the conformal factor relating the static and Bjorken metrics in the next-to-next-to-leading order we may obtain the temperature as
\be
T = T_H \left( \frac{1}{\tau^{1/3}} - \frac{\mathcal{C}_1}{\tau} +\frac{\mathcal{C}_2}{\tau^{5/3}} \right)
\ee
The Stefan-Boltzmann law may be used to calculate the other thermodynamic quantities in the next-to-next-to-leading order. In particular, the viscosity to entropy density ratio is still given by (\ref{eqetaNLO}) with $d=5$, but with $\mathcal{C}_1$ constrained by (\ref{eqCco}). It follows that this ratio is given by
\be \frac{\eta}{s} = \frac{1}{4\pi} \ee
which is the same value one obtains from sinusoidal perturbations of the AdS Schwarzschild metric \cite{son}. This is also in agreement with the conclusion reached by considering time-dependent asymptotic solutions of the Einstein equations \cite{Janik}.

\section{Conclusion}
\label{sec4}

We discussed the possibility of obtaining viscous Bjorken hydrodynamics on a $(d-1)$-dimensional Miskowski space from a large AdS$_d$ Schwarzschild black hole (of flat horizon).
The latter is normally considered dual to a static gauge theory fluid on the boudary whose temperature coincides with the Hawking temperature.
By appropriately modifying the boundary conditions, we obtained viscous Bjorken hydrodynamics on the boundary in the limit of large longitudinal proper time ($\t \to\infty$) at next-to-leading order. Our results are in agreement with those obtained by considering time-dependent asymptotic solutions of the Einstein equations in five dimensions~\cite{JP,NakSin,SinNakKim}.
Moreover, since our bulk space consisted of a Schwarzschild black hole,
we were able to determine the temperature of the conformal fluid on the boundary in terms of the Hawking temperature of the hole.
At next-to-next-to-leading order, we saw that no choice of boundary conditions could lead to a boost-invariant flow. In order to obtain a dual Bjorken flow at that order, we had to perturb the black hole metric. This led to a constraint on the viscosity coefficient and the viscosity to entropy density ratio was fixed to the value $1/(4\pi)$ as in the case of sinusoidal perturbations \cite{son}. This was in agreement with a next-to-next-to-leading order calculation of a time-dependent asymptotic solution \cite{Janik}.

We should point out that a constraint on the viscosity is not necessary if one does not perturb the black hole metric. In this case, one obtains deviations from Bjorken flow which are a subleading effect at late times. It might be worth exploring the connection of such deviations (coming from a dual Schwarzschild black hole) to existing (RHIC) and forthcoming (LHC) experimental data.

It may also be worthwhile, albeit tedious, to go beyond the perturbative order considered here. It has already been observed that the supergravity Fefferman-Graham metrics dual  
to boost-invariant hydrodynamics suffer from singularities of the  
curvature invariants near the reputed black hole horizon \cite{SUGRA}.  This occurs at the third perturbative order it the large $\tau$ expansion  
and cannot be canceled by an appropriate choice of transport coefficients as has been done at second order.  However, by working with  
Eddington-Finkelstein instead of Fefferman-Graham coordinates, an expansion in the new time coordinate was shown to lead to non-singular solutions at all orders with the correct choice of transport  
coefficients \cite{BBB1,BBB}.  It would be interesting to investigate the connection of the AdS Schwarzschild black hole metric with these Eddington-Finkelstein-type solutions of the Einstein equations.

\section*{Acknowledgment}
Work supported in part by the Department of Energy under grant DE-FG05-91ER40627.

\end{document}